\documentclass{article}

\usepackage[dblblindworkshop, final]{neurips_2025}

\workshoptitle{ML for the Physical Sciences}

\usepackage[utf8]{inputenc} 
\usepackage[T1]{fontenc}    
\usepackage{hyperref}       
\usepackage{url}            
\usepackage{booktabs}       
\usepackage{amsfonts}       
\usepackage{nicefrac}       
\usepackage{microtype}      
\usepackage{xcolor}         
\usepackage{graphicx}
\usepackage{amsmath}        
\usepackage{nth}
\title{Extrapolating Phase-Field Simulations in Space and Time with Purely Convolutional Architectures}

\author{Christophe Bonneville\thanks{Sandia National Laboratories, Livermore, CA 94550; \texttt{\{cpbonne,probbe,hnnajm,csafta\}@sandia.gov}} 
\and \textbf{Nathan Bieberdorf}\thanks{Materials Sciences Division, Lawrence Berkeley National Laboratory \texttt{and} Department of Materials Science and Engineering, University of California, Berkeley, CA 94720, USA; \texttt{\{nbieberdorf,mdasta\}@berkeley.edu}} 
\and  \textbf{Pieterjan Robbe}\footnotemark[1]
\and  \textbf{Mark Asta}\footnotemark[2]
\and  \textbf{Habib N. Najm}\footnotemark[1]
\and  \textbf{Laurent Capolungo}\thanks{Los Alamos National Laboratory, Los Alamos, NM 87544; \texttt{laurent@lanl.gov}}
\and  \textbf{Cosmin Safta}\footnotemark[1]
}

\begin{document}

\maketitle

\begin{abstract}
Phase-field models of liquid metal dealloying (LMD) can resolve rich microstructural dynamics but become intractable for large domains or long time horizons. We present a conditionally parameterized, fully convolutional U-Net surrogate that generalizes far beyond its training window in both space and time. The design integrates convolutional self-attention and physics-aware padding, while parameter conditioning enables variable time-step skipping and adaptation to diverse alloy systems. Although trained only on short, small-scale simulations, the surrogate exploits the translational invariance of convolutions to extend predictions to much longer horizons than traditional solvers. It accurately reproduces key LMD physics, with relative errors typically under 5$\%$ within the training regime and below 10$\%$ when extrapolating to larger domains and later times. The method accelerates computations by up to 16,000 times, cutting weeks of simulation down to seconds, and marks an early step toward scalable, high-fidelity extrapolation of LMD phase-field models.
\end{abstract}

\section{Introduction}

Machine learning has become a powerful tool for accelerating simulations governed by partial differential equations (PDEs), enabling surrogates that run orders of magnitude faster than traditional solvers \cite{karniadakis2021physicsinformed,raissi2017physicsinformeddeeplearning,li2021fourier,lu2021deeponet,kovachki2021neural,BONNEVILLE2024116535,bonneville2024comprehensive}. Much prior work, however, targets \textit{simplified problems} where large training datasets are reasonably cheap to generate. In such cases, models are trained on complete simulations and rarely tested under conditions far beyond the training data.

Here, we investigate phase-field simulations of liquid-metal de-alloying (LMD) \cite{geslin2015,Lai:2022a, Lai:2022, bieberdorf2023,KerrBieberdorf2024,Bieberdorf2025}, where creating an ideal dataset would require years of high-performance computing time. Phase field simulations for LMD are governed by physically-derived PDEs and can be used to model microstructure and morphology evolution of material systems \cite{LQChenReview,KarmaReview,SteinbachReview}. We focus on binary alloys in contact with a corrosive liquid, and use a thermokinetic parameterization that triggers rapid selective dissolution from the binary alloy and leads to formation of fine-scale, porous microstructures. The de-alloyed microstructures exhibit a wide variety of nano-scale morphological patterns that ultimately control macro-scale properties, such as material strength. While traditional phase-field solvers can simulate LMD \cite{geslin2015,Lai:2022a,Lai:2022,bieberdorf2023,KerrBieberdorf2024,Bieberdorf2025}, they are limited to short durations and small domains, preventing direct numerical simulations (DNS) from reaching the spatial and temporal scales required in practice.

This scarcity of data is problematic for common machine learning approaches to PDE forecasting, particularly auto-regressive methods \cite{kovachki2021neural,li2021fourier,Lu_2021,cao2023lno,raonić2023convolutional}, which evolve systems step by step, feeding predictions back as inputs. These methods tend to accumulate errors over long horizons. In simple problems, abundant long sequences help mitigate instability, but for LMD only short sequences are available, and models must remain stable when extrapolated far beyond training. Prior LMD surrogates, such as U-AFNO \cite{bonneville_2025}, while promising, lacked this extrapolation capability.

To overcome these challenges, we design a modified U-Net \cite{ronneberger2015u} surrogate that exploits the translation invariance of convolutions to handle variable domain sizes while preserving physical consistency. The architecture uses parametric conditioning \cite{oommen2023rethinking} to adapt to different alloy and simulation parameters, and circular padding to strictly enforce boundary conditions. This design ensures stability in long auto-regressive roll-outs and scales to domains and horizons much larger than those seen in training. Our model is trained on short simulations over square domains, but during surrogate predictions, it dynamically extends domain height as corrosion progresses and the corrosive bath sinks into the alloy, enabling extrapolation in both space and time.

\section{Phase-Field Model for Liquid Metal De-alloying}

The phase-field model simulates LMD in a binary alloy \cite{geslin2015,bieberdorf2023}, resolving both the diffuse solid-liquid interface $\phi$ and the mole fractions of alloy species $c_\mathrm{A}$, $c_\mathrm{B}$, and $c_\mathrm{C}$ (A and B are part of the alloy, and C is the corrosive agent. Since $c_C=1-c_A-c_B$, we only track A and B). The interface and conserved species dynamics are governed by coupled modified Allen-Cahn and Cahn-Hilliard equations:
\begin{equation}
\label{phasefield}
\frac{\partial \phi}{\partial t} = -\tilde{M}_\phi \frac{\pi^2}{8 \eta} \frac{\delta F}{\delta \phi} 
\hspace{0.5in}
\frac{\partial c_i}{\partial t} = \nabla \cdot \sum_{j=\mathrm{A,B}} M_{ij}(\phi) \nabla \left( \frac{\delta F}{\delta c_j} \right), \quad i \in \{\mathrm{A,B}\}
\end{equation}
where $F$ is the free energy functional containing both interface and bulk chemical contributions. Ultimately, equation \ref{phasefield} relies on fourth-order spatial derivative terms, and is numerically stiff. Conventional time-integrators require an extremely small time step of $\Delta t = 10^{-12}\,\mathrm{s}$ to maintain stability \cite{bieberdorf2023,KerrBieberdorf2024}. This makes direct simulation over realistic timescales computationally infeasible, motivating the use of auto-regressive surrogate models that can leap over many solver steps at once, while preserving the ability to predict long-term system evolution.

\section{U-Net Surrogate}

\begin{figure}[!ht]
\centering
    \includegraphics[width=1.\textwidth]{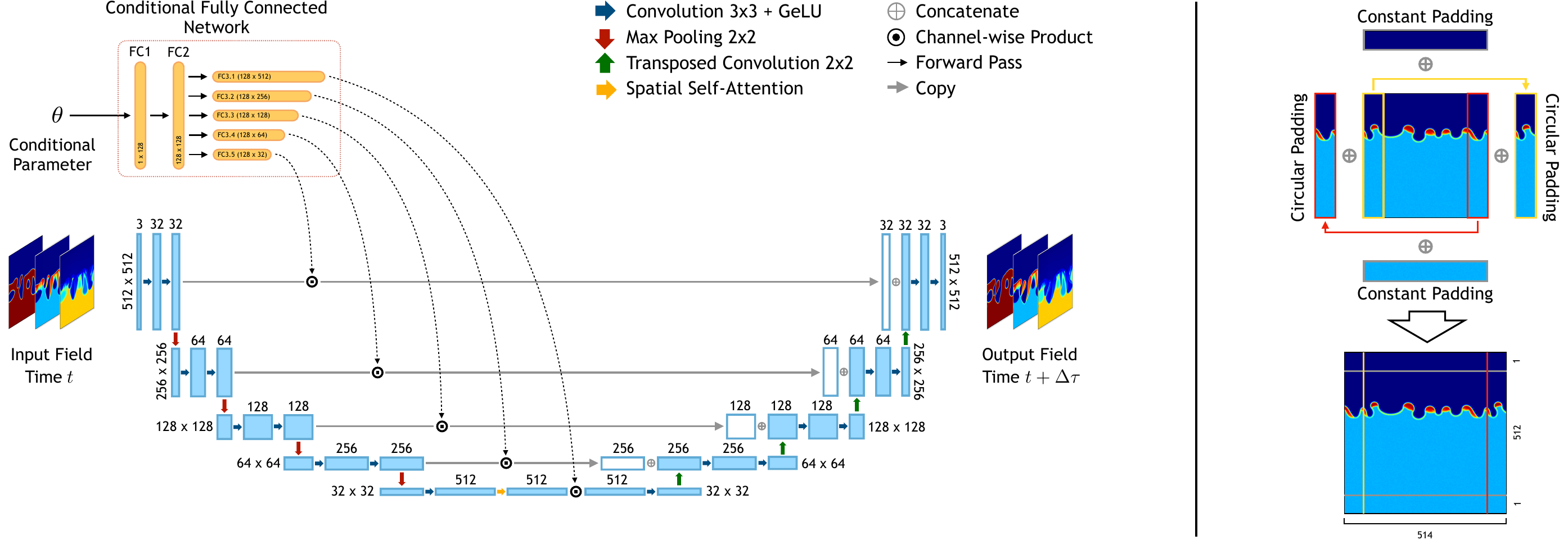}
    \caption{Left: U-Net Architecture. The model takes as input the field at time $t$ and outputs the field at a future time step $t+\Delta\tau$ and is conditioned on $\theta$. Right: Circular padding for the U-Net convolutions (widths of the padded vectors are exaggerated for illustration).}
    \label{fig:cunet}
\end{figure}

U-Nets \cite{ronneberger2015u} are residual convolutional neural networks with a U-shaped encoder–decoder structure linked by skip connections, enabling efficient gradient propagation and accurate tensor-to-tensor regression. Originally developed for biomedical segmentation, they are now widely used for PDE-based physical simulations, particularly in autoregressive prediction where they map spatial input fields to future states \cite{ovadia2023vito,liu2023ditto,oommen2023rethinking,oommen_deeponet}. While not resolution invariant like neural operators \cite{li2021fourier,lu2021deeponet}, their convolutional locality allows training on small domains and applying to larger ones at fixed resolution. Outputs are passed through a sigmoid to enforce the [0,1] bounds of the phase and species fields. We use a conditional U-Net inspired by \cite{oommen2023rethinking}, where parameters modulate the network via a fully connected conditioning module. The conditioning inputs are the time span between U-Net input and output, and the reference concentration $c_A$. These pass through two hidden layers of 128 neurons with SiLU activation, followed by five linear layers producing scaling vectors of sizes 32, 64, 128, 256, and 512. Vectors are applied to the corresponding U-Net feature channels before skip concatenation, enabling multi-step predictions and adaptation to varying material conditions without retraining. The U-Net architecture is shown in figure \ref{fig:cunet}

We integrate spatial self-attention in the bottleneck to capture long-range dependencies, building on prior work that used Vision Transformers (ViTs) in U-Net bottlenecks \cite{bonneville_2025, ovadia2023realtime, ovadia2023vito}. Rather than adopting the full transformer architecture (with fixed-size linear layers constraining input–output dimensions), we retain only the attention layers \cite{bahdanau2015neural}, implemented entirely with convolutions to preserve resolution flexibility and allow application to domains larger than those seen in training. Queries, keys, and values are computed with $1\times1$ convolutions, with query/key channels set to $1/8^\mathrm{th}$ of the input dimension. The U-Net convolutional layers employ $3\times3$ kernels, so padding is needed to preserve spatial dimensions. Instead of standard zero-padding, which ignores physics, we design boundary-aware padding \cite{alguacil2021effectsboundaryconditionsfully,Ren_2022}. Horizontally, we use circular padding, which hard-enforces periodic boundary conditions by wrapping left and right edges onto each other. At the top we apply zero-padding, since both phase and species fields vanish there, and at the bottom we replicate boundary values to reflect constant composition. This embeds known boundary behaviors directly into the network, improving accuracy and generalization. 

\section{Training and Testing}

The training dataset contains simulations with reference $c_A$ concentrations from 0.20 to 0.40 in 0.05 increments (1,280 snapshots per concentration). Data is generated on a 512$\times$512 grid, depicting species C sinking through the A–B alloy toward the lower boundary. Besides $c_A$, the model is designed to skip 50,000–100,000 time steps per iteration and is implemented in PyTorch \cite{paszke2019pytorch}. It is trained with relative $\mathcal{L}_2$ loss (common in neural-operator models \cite{li2021fourier, lu2021deeponet, bonneville_2025}), Adam optimization \cite{kingma2017adam}, and a $10^{-4}$ learning rate, for 80 epochs. Training on a single A100 GPU requires 59 hours. Surrogate tests use an auto-regressive roll-out, recursively feeding predictions back as inputs \cite{bonneville_2025, oommen2023rethinking}. Leveraging U-Net’s convolutional design, we run extended domains twice as tall as training (1024$\times$512) to capture deeper species C penetration, and extrapolate to horizons about 3$\times$ longer than in training. This domain size is chosen as the largest possible in the limit of computational feasibility, since generating DNS test data at 1024$\times$512 can take weeks/months on 64–128 CPU cores.

Because the physics is highly chaotic \cite{bieberdorf2023}, the surrogate cannot reproduce exact microstructures. Instead, the goal is to capture invariant physical trends across simulations \cite{bonneville_2025,oommen2023rethinking,kovachki2021neural}. Accuracy is defined by recovering key quantities of interest (QoIs) extracted from predicted fields rather than the fields themselves. QoIs include interface curvature and perimeter, C penetration depth, remaining A and B volumes, and average ligament height \cite{McCue_Karma_Erlebacher_2018,Wada:2023,VAKILI20171852,Tran_2019,Lai:2022,MCCUE201610,bonneville_2025}. Surrogate QoIs are compared against DNS using relative $\mathcal{L}_2$ error.

\section{Results}

\begin{figure}[!ht]
\centering
    \includegraphics[width=1.\textwidth]{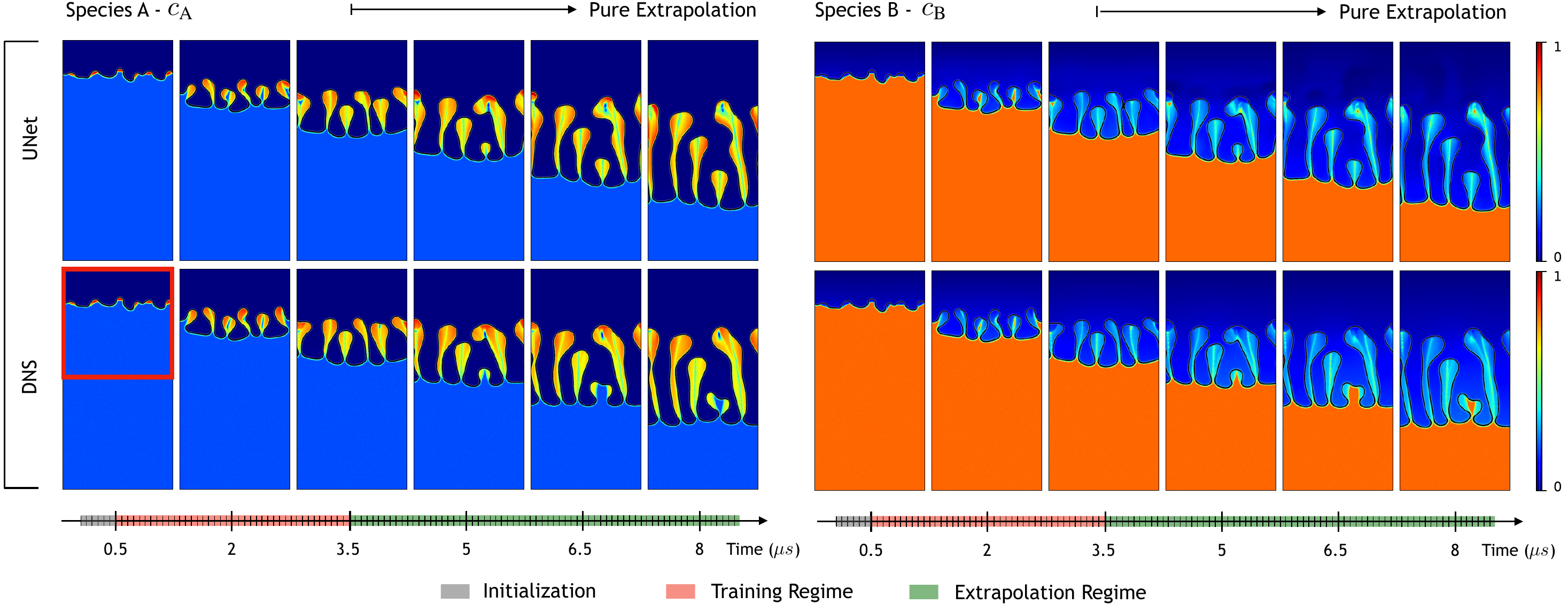}
    \caption{Concentration fields for species A and B. The reference (pre-dealloying) concentration for A is $\pmb{c_A=0.2}$. For each species, the top row represents the U-Net surrogate prediction, and the lower row is the corresponding direct numerical simulation (DNS). The red box illustrates the size of the domain used in the training data.}
    \label{fig:field_020}
\end{figure}

Figure \ref{fig:field_020} shows test simulations for $c_A=0.2$, comparing U-Net predictions with the corresponding DNS. Both sets of fields illustrate the evolution of species A and B, with the phase-field interface $\phi$ outlined in black. Note that the surrogate simulations start at $t=0.5\mu$s. This is because at $t=0\mu$s, the initial interface is perfectly flat. Such an initial condition is not directly suitable for the U-Net, so we initially run the numerical solver for $0.5\mu$s to initialize the surrogate. Because the system is highly chaotic, the predicted microstructures inevitably diverge in detail from the true ones over long time horizons. Nevertheless, the surrogate reproduces the key physics: species C penetrates the alloy at the correct pace, the interface shifts downward realistically, and the curvature of the interface aligns closely with that of the true fields. The ligament structures also form and evolve in a physically consistent way, including the characteristic spikes in species A near the interface and the erosion of species B. Importantly, the surrogate remains numerically stable, producing no unphysical artifacts, even deep into the extrapolation regime. Predictions continue to unroll smoothly, preserving the correct physical trends well beyond the training window. Note that here, \textit{training regime} only refers to the time span used in the training data, while the surrogate is being compared with completely new and independent testing data. 

\begin{figure}[!ht]
\centering
    \includegraphics[width=1\textwidth]{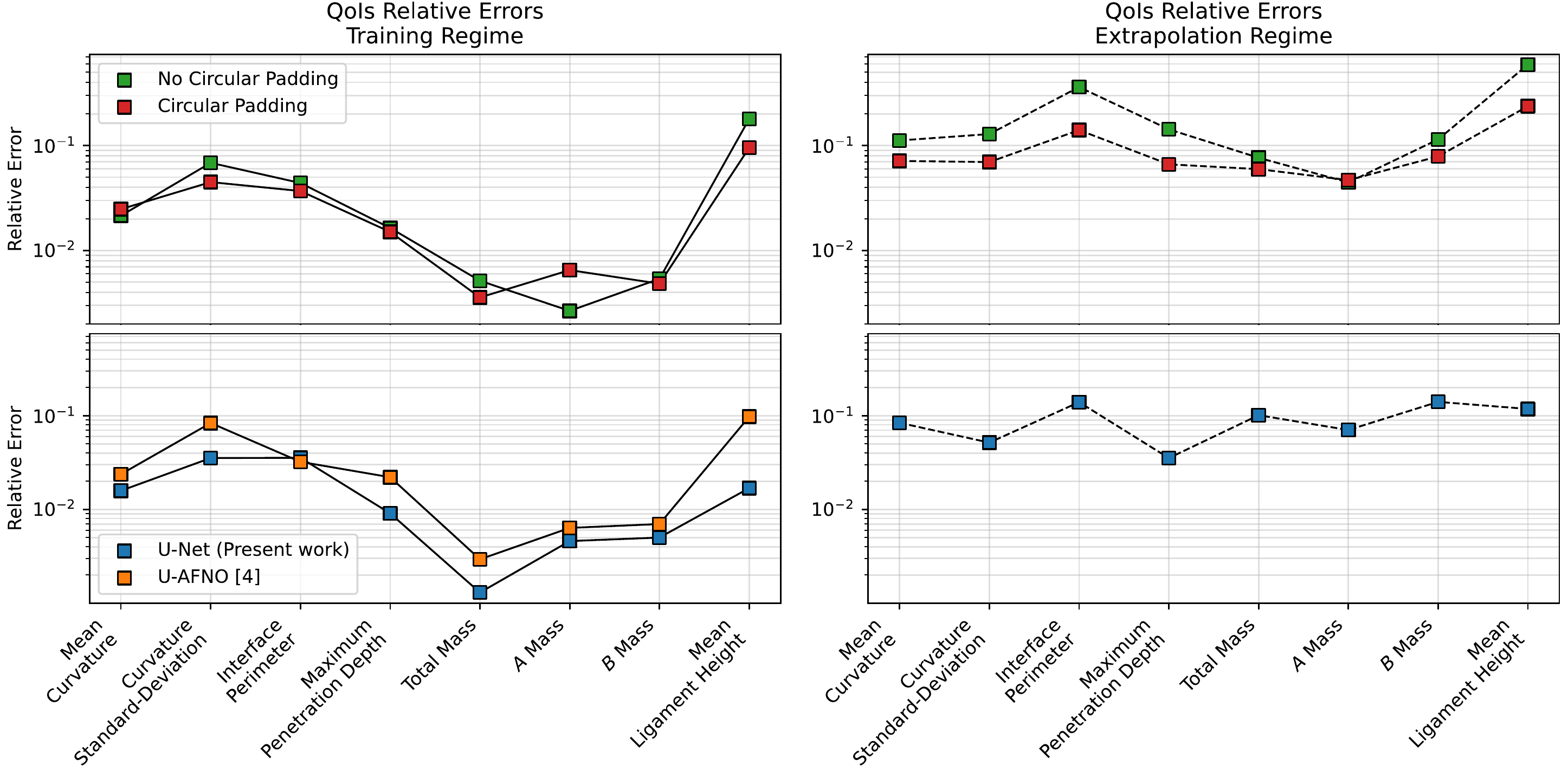}
    \caption{QoIs relative errors. Top row: Errors for the U-Net with and without circular padding (averaged across all tested concentration for $c_A\in[0.2,0.4]$. Bottom row: The present model (U-Net) performance vs. U-AFNO \cite{bonneville_2025}, for $c_A=0.3$ (note that the U-AFNO cannot extrapolate).}
    \label{fig:unet_vs_uafno}
\end{figure}

Figure \ref{fig:unet_vs_uafno} reports QoI relative errors in both the training and extrapolation regimes, comparing surrogate performance with and without circular padding. Errors in the training regime remain below 5\%, and even during extrapolation it typically stays within 10\%. The mean ligament height proves the most difficult QoI to predict. Removing periodic convolutions tends to worsen accuracy, particularly in extrapolation, as mismatched lateral boundaries introduce instabilities and unphysical artifacts in the QoIs. We also benchmark our surrogate against U-AFNOs, a recent surrogate for LMD simulations presented in Bonneville et al. \cite{bonneville_2025}. In this case, the comparison is restricted to $c_A=0.3$ and within the training regime, since the earlier approach in \cite{bonneville_2025} cannot handle other concentrations or domain extrapolation. Overall, our surrogate achieves up to 5\% better accuracy across most QoIs.

In conclusion, our surrogate demonstrates the ability to extrapolate at least three times beyond the training horizon, enabled by its fully convolutional design and robust enforcement of boundary conditions. The model delivers both high accuracy and long-term stability. For context, a reference simulation such as the one shown in Figure 2 requires approximately 7.5 days of computation, whereas our surrogate reaches the same time horizon in just 38.5 seconds on a single A100 GPU, a speed up of about 16,000 times.

\begin{ack}
This work was supported by the U.S. Department of Energy, Office of Nuclear Energy, and Office of Science, Office of Advanced Scientific Computing Research through the Scientific Discovery through Advanced Computing project on Simulation of the Response of Structural Metals in Molten Salt Environment.
This article has been co-authored by employees of National Technology and Engineering Solutions of Sandia, LLC under Contract No. DE-NA0003525 with the U.S. Department of Energy (DOE). The employees co-own right, title and interest in and to the article and are responsible for its contents. The United States Government retains and the publisher, by accepting the article for publication, acknowledges that the United States Government retains a non-exclusive, paid-up, irrevocable, world-wide license to publish or reproduce the published form of this article or allow others to do so, for United States Government purposes. The DOE will provide public access to these results of federally sponsored research in accordance with the DOE Public Access Plan \url{https://www.energy.gov/downloads/doe-public-access-plan}. Sandia Release Number: SAND2025-10690O.
Los Alamos National Laboratory, United States, an affirmative action/equal opportunity employer, is operated by Triad National Security, LLC, for the National Nuclear Security Administration of the U.S. Department of Energy under Contract No. 89233218CNA000001.
Lawrence Berkeley National Laboratory is supported by the DOE Office of Science under contract no. DE-AC02-05CH11231. This study made use of computational resources of the National Energy Research Scientific Computing Center (NERSC), which is also supported by the Office of Basic Energy Sciences of the US Department of Energy under the same contract number.
\end{ack}

\bibliography{references}

\end{document}